\documentclass[11pt]{article}
\usepackage[letterpaper,margin=0.86in,headheight=14pt,headsep=17pt,footskip=27pt]{geometry}
\usepackage[T1]{fontenc}
\usepackage[utf8]{inputenc}

\usepackage{polyglossia}
\setdefaultlanguage{english}
\setotherlanguage{russian}
\newfontfamily\russianfont{DejaVuSerif.ttf}[
  BoldFont=DejaVuSerif-Bold.ttf,
  ItalicFont=DejaVuSerif-Italic.ttf,
  BoldItalicFont=DejaVuSerif-BoldItalic.ttf,
  Scale=0.95]
\usepackage{amsmath,amssymb}
\usepackage{graphicx}
\usepackage{microtype}
\usepackage{fancyhdr}
\usepackage{enumitem}
\usepackage{xcolor}
\usepackage{hyperref}
\hypersetup{
  unicode=true,
  colorlinks=true,
  linkcolor=black,
  citecolor=black,
  urlcolor=black,
  pdftitle={On the Theory of Multidimensional Stationary Random Processes (1941): Translation and Commentary},
  pdfauthor={V. N. Zasukhin (original author); Bulat Gafarov and Madina Karamysheva (translation and commentary)},
  pdfsubject={Historical translation and commentary; multivariate Wold decomposition; stationary random processes},
  pdfkeywords={Zasukhin, Zasuhin, Wold decomposition, multivariate prediction, stationary processes}
}
\newcommand{\M}{M}
\newcommand{\refsup}[1]{\textsuperscript{(#1)}}
\newcommand{\origpage}[1]{\par\medskip\noindent{\small\textbf{[#1]}}\par\smallskip}
\newcommand{\deflabel}[1]{\par\medskip\noindent\textbf{#1.}}
\newcommand{\thmlabel}[1]{\par\medskip\noindent\textbf{#1.}}
\newcommand{\sectiontitle}[1]{\par\bigskip\noindent{\large\bfseries #1}\par\medskip}
\begin{document}
\pdfbookmark[0]{Title and abstract}{edition-title}
\thispagestyle{plain}
\begin{center}
{\large V. N. Zasukhin (1941)\par}
\vspace{8pt}
{\LARGE\bfseries On the Theory of Multidimensional\\[3pt]
Stationary Random Processes\par}
\vspace{6pt}
{\large\bfseries Translation and Commentary\par}
\vspace{17pt}
{\normalsize Original author: \textbf{V. N. Zasukhin (1915--1941)}\par}
\vspace{8pt}
{\normalsize English translation and commentary by\par}
{\large\bfseries Bulat Gafarov and Madina Karamysheva\par}
{\small Bulat Gafarov: Department of Agricultural and Resource Economics\\
University of California, Davis\par}
{\small Madina Karamysheva: HSE University\par}
\vspace{10pt}
{\small October 26, 2026\par}
\end{center}
\vspace{6pt}

\begin{abstract}
This edition provides an English translation and commentary on
V. N. Zasukhin's 1941 note
\textquotedblleft On the Theory of Multidimensional Stationary Random
Processes,\textquotedblright\ originally published in
\textit{Doklady Akademii Nauk SSSR}, 33(7--8), 435--437.
The note states a multivariate extension of Wold's decomposition: an
orthogonal decomposition into singular and regular components, a one-sided
moving-average representation of the regular component in terms of
orthonormal innovation sequences, and a spectral criterion for regularity
of full rank. The latter combines absolute continuity of the spectral
functions with positivity and logarithmic integrability of the determinant
of the spectral density matrix. Whittle (1953) explicitly credited Zasukhin
with the multivariate generalization. The original results are presented
without detailed proofs. A separate editorial note discusses their
historical reception and explains the terminology. A retypeset Russian
transcription is included for comparison.
\end{abstract}

\sectiontitle{Source and editorial conventions}
\noindent\textbf{Original publication.}
V. N. Zasukhin, \textit{\textrussian{К теории многомерных стационарных
случайных процессов}}, \textit{Doklady Akademii Nauk SSSR},
vol.~33, nos.~7--8 (1941), pp.~435--437.
Presented by Academician A. N. Kolmogorov on November 10, 1941.

The English text retains the original definitions, theorem and equation
numbers, and four-item bibliography. Bracketed labels [435]--[437] indicate
the original pagination; the page numbers of this edition are consecutive.
The Russian appendix is a transcription, not a photographic facsimile.
The abstract and the following editorial note are additions to the 1941
paper and are not attributed to Zasukhin.

Typographical normalization in the English version includes writing the
summation limit \(j=q\) as \(q\), and using \(\{s(t)\}\) and \(\{r(t)\}\)
for the component-indexed process notation in Theorem~1. Neither change
is presented as a correction of the original mathematics. The Russian
transcription preserves these forms.

\clearpage
\pdfbookmark[0]{Editorial note}{editorial-note}
\sectiontitle{Editorial note}
\begingroup
\small
\linespread{0.92}\selectfont

\noindent\textbf{Zasukhin's contribution and its reception.}
The note extends Wold's one-dimensional decomposition to vector-valued
stationary processes. Theorems~1 and~2 give the orthogonal
singular--regular decomposition and the one-sided innovations
representation, allowing the innovation rank \(q\) to be smaller than
the number \(n\) of observed coordinates. Theorem~3 states the spectral
criterion for regularity of full rank.  

Whittle's attribution of multivariate Wold extention to Zasukhin is explicit. In his paper on multiple stationary
time series, Whittle \cite[p.~127]{Whittle1953} writes:
\begin{quote}
\textquotedblleft Zasuhin (1941) has generalized Wold's representation
to the multivariate case.\textquotedblright
\end{quote}
Thus Whittle credits the earlier multivariate generalization to Zasukhin.
The spelling \textit{Zasuhin}, used by Whittle and by Wiener and Masani,
is a transliteration variant of \textit{Zasukhin}.

Wiener and Masani \cite[pp.~111--112, 136]{WienerMasani1957}
likewise trace the development along Kolmogorov's lines to Zasukhin's
1941 work. They identify the multivariate Wold decomposition with
his Theorem~1, and the full-rank regularity criterion with his
Theorem~3. They also emphasize the absence of proofs. Zasukhin himself
indicates that Theorems~1 and~2 can be obtained by Kolmogorov's methods;
no detailed proofs are supplied in the note.  

Wiener and Masani \cite[p.~111, footnote~1]{WienerMasani1957}
acknowledge G. Kallianpur for an English translation of Zasukhin's paper
during their 1955--1956 collaboration. An earlier English translation
therefore existed; this edition makes no claim to be the first one.

\medskip
\noindent\textbf{Innovation rotations and structural identification.}
The last assertion of Theorem~2 is also noteworthy: the innovation rank
\(q\) is uniquely determined, but two normalized innovation representations
of the same regular process differ by a \emph{time-invariant} change of
orthonormal coordinates. In modern matrix notation, if
\(r_t=\sum_{j\geq 0} C_j u_{t-j}\), the same representation can be written
with \(u_t^{\star}=Q u_t\) and \(C_j^{\star}=C_j Q^{*}\), where
\(Q^{*}Q=I_q\). For real-valued processes the admissible changes of basis
form the orthogonal group \(O(q)\); when \(q\geq 2\), its continuous
dimension is \(q(q-1)/2\). Theorem~2 thereby identifies an exact source
of nonuniqueness in multivariate linear prediction, without requiring
Gaussianity or assigning a causal meaning to the innovation coordinates.

For an unrestricted Gaussian structural VAR, this becomes the familiar
observational equivalence of orthogonally rotated structural shocks;
see also Leeper, Sims, and Zha \cite[pp.~4--5]{LeeperSimsZha1996}
and Rubio-Ram\'{i}rez, Waggoner, and Zha
\cite[Theorem~1]{RubioRamirezWaggonerZha2010}.
Leeper, Sims, and Zha explicitly exhibit the orthogonal transformation
and its \(n(n-1)/2\) free parameters. Rubio-Ram\'{i}rez, Waggoner, and
Zha characterize equivalent structural parameterizations and, following
Rothenberg \cite{Rothenberg1971}, distinguish global from local
identification. The necessary \emph{order condition} requires
\(n(n-1)/2\) restrictions; additional conditions determine whether
the restrictions actually identify the model. Read retrospectively,
Zasukhin's 1941 observation anticipates this \emph{mathematical
indeterminacy}, preceding the econometric treatments of simultaneous
equations and stochastic difference equations by Haavelmo
\cite{Haavelmo1943} and Mann and Wald \cite{MannWald1943}, as well as
Sims's structural-VAR discussion \cite{Sims1980}. 

\newpage
\noindent\textbf{Zasukhin's death in 1941.}
Kolmogorov's memorial \cite{Kolmogorov1970} records Zasukhin's life
dates, 1915--1941, and states that he died at the front in 1941.
Tyulina's subsequent memorial history of Moscow University mathematicians
and mechanicians \cite{Tyulina2009} gives a more specific account.
According to letters Zasukhin sent to his family and information from
surviving comrades, he served as an infantry lieutenant in a division
defending Mozhaisk, was subsequently transferred to the vicinity of
Yelnya (Smolensk region), and was killed there in October 1941.
Tyulina's account identifies the approximate place and month, but does
not establish the exact date, village, or burial site.
\footnote{A portrait and biographical sketch appear in Odinets
\cite[pp.~59--61; portrait on p.~59]{Odinets2024}.
Odinets reports that Zasukhin went missing in action in September 1941;
his dating differs from Tyulina's October account.}

\begin{figure}[htbp]
\centering
\includegraphics[width=0.31\textwidth]{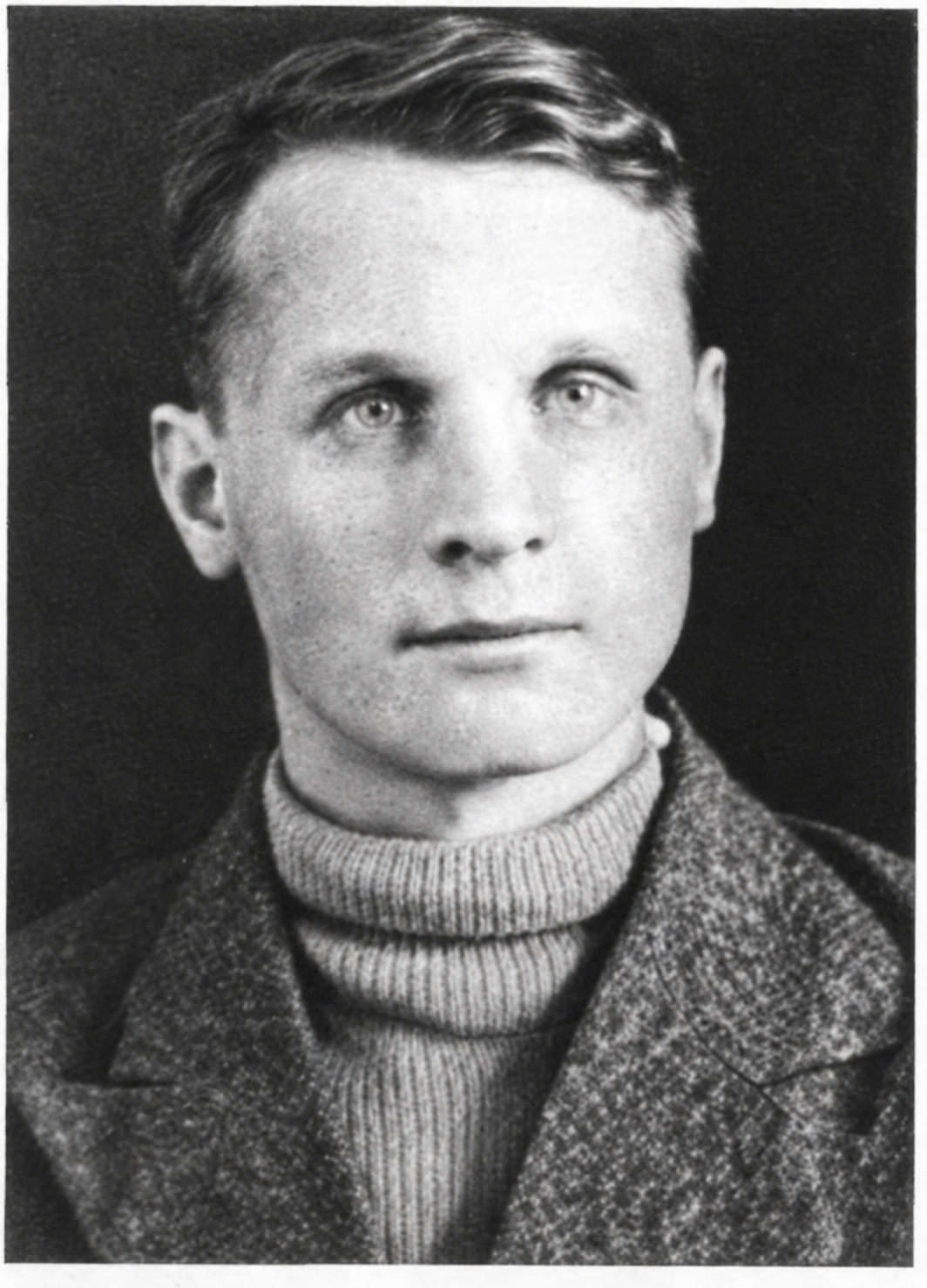}
\caption{Viktor Nikolaevich Zasukhin (1915--1941).
Historical photograph reproduced from Tyulina (2009),
\emph{In Memory of the Mathematicians and Mechanicians of Moscow
University Who Perished in the Great Patriotic War}.
Original photographer unidentified.}
\label{fig:zasukhin-portrait}
\end{figure}

\medskip
\noindent\textbf{Terminology and mathematical conventions.}
The historical \textit{unitary space} is the complex \(L^2\) Hilbert-space
setting. For the modern projection interpretation, the spans \(H_x\)
and \(H_x(t)\) are understood as closed linear spans, and the infinite
series in Theorem~2 converges in mean square. The original text is not
silently rewritten to insert these conventions.
\textit{Singular} here means linearly deterministic, not merely a singular
spectral density matrix. \textit{Regular} corresponds to pure nondeterminism,
with the zero process excluded by Definition~2.
The assumptions use raw second moments and do not impose zero means.
Orthogonality of the innovations does not assert independence or
Gaussianity. Theorem~2 retains the source notation \(H_x(t)\).

\renewcommand{\refname}{Editorial references}
\begingroup
\footnotesize
\setlength{\parskip}{0pt}

\endgroup
\endgroup

\clearpage
\pdfbookmark[0]{English translation}{english-translation}
\fancyhead[R]{\footnotesize English translation}
\setcounter{equation}{0}
\renewcommand{\theHequation}{en.\arabic{equation}}
\begin{center}
{\small\bfseries ENGLISH TRANSLATION\par}
\vspace{8pt}
{\Large\bfseries On the Theory of Multidimensional\\
Stationary Random Processes\par}
\vspace{8pt}
{\large V. N. Zasukhin\par}
\vspace{6pt}
{\itshape (Presented by Academician A. N. Kolmogorov on 10 XI 1941)\par}
\end{center}
\origpage{435}
Let a complex random variable $x_k(t)$ correspond to each integer
$t\,(-\infty<t<+\infty)$ and each $k=1,2,\ldots,n$. The collection of random
variables $x_k(t)$ forms an $n$-dimensional discrete stationary random process
in the sense of H. Cram\'er (see \refsup{1}) if the expectations
\[
B_{ij}(m)=\M\bigl[x_i(t+m)\overline{x_j(t)}\bigr]
\]
are finite and do not depend on $t$ (the additional assumption
$\M[x_i(t)]=0$ adopted in H. Cram\'er's paper is immaterial for what follows).
Henceforth, a collection of random variables of this kind will simply be
called an $n$-dimensional process and denoted by $\{x(t)\}$.

As indicated in the papers of A. N. Kolmogorov \refsup{2} and \refsup{3}, in
studying stationary random processes it is natural to use the geometry of the
unitary space $\mathfrak S$ of all complex random variables $z$ in the problem
having finite expectation $\M|z|^2$. Following A. N. Kolmogorov, denote by
$H_x$ the smallest linear subspace of $\mathfrak S$ containing all elements
$x_k(t)$, $-\infty<t<+\infty$, $k=1,2,\ldots,n$, and by $H_x(t)$ the smallest
linear subspace of $\mathfrak S$ containing all $x_k(s)$ for which $s\le t$.

\deflabel{Definition 1} \emph{The process $\{x(t)\}$ is called singular if
$H_x(t)=H_x$ for every $t$.}

Singular processes coincide with processes that permit an arbitrarily accurate
prediction of their future course (the values $x_k(t)$ for $t>t_0$) from their
course in the past (the values $x_k(t)$ for $t\le t_0$) by means of linear
prediction formulas with constant coefficients (see, in this connection, the
paper of A. N. Kolmogorov \refsup{2}).

\deflabel{Definition 2} \emph{The process $\{x(t)\}$ is called regular if the
intersection $S_x$ of the spaces $H_x(t)$ corresponding to all integer values
of $t$ consists of the single point $0$, but not all $x_k(t)$ are zero.}

The following theorem shows that an arbitrary nonsingular 

\origpage{436}

process can be
represented in a completely definite way as the sum of a singular and a
regular process.

\thmlabel{Theorem 1} \emph{Every $n$-dimensional nonsingular process
$\{x(t)\}$ can be represented in the form}
\begin{equation}
 x_k(t)=s_k(t)+r_k(t),
\end{equation}
\emph{where the $n$-dimensional processes $\{s(t)\}$ and $\{r(t)\}$ have the
following properties:}

\noindent\emph{1) the process $\{s(t)\}$ is singular;}\par
\noindent\emph{2) the process $\{r(t)\}$ is regular;}\par
\noindent\emph{3) $\M[s_i(s)\overline{r_j(t)}]=0$ for all $i,j,s$, and $t$;}\par
\noindent\emph{4) for every $k$ and $t$, the elements $s_k(t)$ and $r_k(t)$
belong to the space $H_x$.}

\emph{The representation (1) satisfying conditions 1)--4) is unique.}

The processes $\{s(t)\}$ and $\{r(t)\}$ satisfying the conditions of Theorem 1
are called, respectively, the singular and regular components of the process
$\{x(t)\}$. The elements $s_k(t)$ of the singular component are precisely the
orthogonal projections of the corresponding elements $x_k(t)$ onto the space
$S_x$. Clearly, a regular process coincides with its regular component. The
nature of regular processes is characterized to a considerable extent by the
following theorem.

\thmlabel{Theorem 2} \emph{Every $n$-dimensional regular process $\{r(t)\}$ can
be represented in the form}
\begin{equation}
 r_k(t)=\sum_{m=0}^{\infty}\sum_{j=1}^{q}c_{kj}(m)u_j(t-m),
\end{equation}
\emph{where:}

\noindent\emph{1) $q\le n$;}\par
\noindent\emph{2) the complex coefficients $c_{kj}(m)$ are such that the matrix
$\|c_{kj}(0)\|$ has rank $q$;}\par
\noindent\emph{3) the random variables $u_k(t)$, $-\infty<t<+\infty$,
$k=1,2,\ldots,q$, form a $q$-dimensional process for which}
\[
\M|u_k(t)|^2=1
\]
\emph{for every $k$ and $t$, and}
\[
\M\bigl[u_i(s)\overline{u_j(t)}\bigr]=0,
\]
\emph{when $i\ne j$ or $s\ne t$;}\par
\noindent\emph{4) the elements $u_k(t)$ belong to the corresponding spaces
$H_x(t)$.}

In all representations of the form (2) of a given regular process satisfying
conditions 1)--4), $q$ has one and the same value. Any two such representations
are related by
\[
 u'_i(t)=\sum_{j=1}^{q}a_{ij}u_j(t),
\]
where the coefficients $a_{ij}$ do not depend on $t$.

The number $q$ is called the rank of the regular process $\{r(t)\}$. The rank
of an arbitrary nonsingular process is defined as the rank of its regular
component. By definition, a singular process

\origpage{437}

has rank zero. Representations
(1) and (2) generalize to $n$-dimensional processes the principal result
obtained by H. Wold in the study of one-dimensional processes (see H. Wold's
book \refsup{4}, or Theorem 18 in \refsup{3}). The proofs of Theorems 1 and 2
formulated above present no particular difficulty if the methods of
\refsup{3} are used.

More difficult are questions concerning the establishment of spectral
characteristics of the various types of stationary processes. In this
direction the author has obtained the following result, which generalizes
Theorem 22 of \refsup{3}.

\newpage
\thmlabel{Theorem 3} \emph{For an $n$-dimensional process $\{x(t)\}$ to be
regular of rank $n$, it is necessary and sufficient that the following three
conditions hold:}

\noindent\emph{1) all spectral functions $F_{ij}(\lambda)$ of the process
$\{x(t)\}$ must be absolutely continuous;}\par
\noindent\emph{2) the determinant}
\[
\Delta(\lambda)=\bigl|F'_{ij}(\lambda)\bigr|
\]
\emph{formed from the derivatives of the spectral functions must be positive
almost everywhere;}\par
\noindent\emph{3) the integral}
\[
J=\int_{-\pi}^{+\pi}\log\Delta(\lambda)\,d\lambda
\]
\emph{must be finite.}

Here the spectral functions $F_{ij}(\lambda)$ are defined, in accordance with
\refsup{3}, by the formulas
\[
B_{ij}(m)=\int_{-\pi}^{+\pi} e^{im\lambda}\,dF_{ij}(\lambda),
\qquad F_{ij}(-\pi)=0.
\]

\vspace{0.8em}
\noindent
\begin{minipage}[t]{0.58\textwidth}
\small Mathematical Institute\\
Moscow State University
\end{minipage}\hfill
\begin{minipage}[t]{0.38\textwidth}
\small\raggedleft Received\\
10 XI 1941
\end{minipage}

\vspace{1.2em}
\begin{center}
{\bfseries Original references}
\end{center}
\begingroup\small
\noindent\textsuperscript{1} H. Cram\'er, \textit{Annals of Mathematics}, 41,
215--230 (1940). \textsuperscript{2} A. N. Kolmogorov, \textit{Izvestiya of
the Academy of Sciences of the USSR, Mathematical Series}, 5, 3--14 (1941).
\textsuperscript{3} A. N. Kolmogorov, \textit{Bulletin of Moscow State
University, Mathematics}, Vol. 2, No. 6 (1941). \textsuperscript{4} H. Wold,
\textit{A Study in the Analysis of Stationary Time Series}, Uppsala, 1938.

\endgroup

\medskip
\noindent\rule{\linewidth}{0.3pt}\par
\begingroup\small
\noindent\textbf{Editorial annotation: subsequent proofs}
(added to this edition; not part of Zasukhin's 1941 text).\par
\smallskip
\noindent\textbf{Theorem 1.} Wiener and Masani
\cite[Lemma~6.10 and Theorem~6.11, pp.~136--138]{WienerMasani1957}
prove the multivariate Wold decomposition using orthogonal projections;
uniqueness follows from the projection construction.\par
\smallskip
\noindent\textbf{Theorem 2.} Wiener and Masani
\cite[Theorem~6.11(b), Corollary~6.12, and Theorem~6.13,
pp.~137--140]{WienerMasani1957}
establish the innovation representation, including deficient innovation
rank. Zasukhin's additional uniqueness statement for normalized
rank-\(q\) innovations, up to a constant unitary change of basis,
follows from uniqueness of the innovation spaces; Wiener and Masani
do not state this part as a separate numbered theorem.\par
\smallskip
\noindent\textbf{Theorem 3.} Wiener and Masani
\cite[Theorem~7.12, pp.~147--148]{WienerMasani1957}
prove the full-rank regularity criterion, using their
Theorems~7.10 and~7.11.\par
\endgroup

\clearpage
\pdfbookmark[0]{Russian text (transcription)}{russian-text}
\fancyhead[R]{\footnotesize Russian text (transcription)}
\setcounter{equation}{0}
\renewcommand{\theHequation}{ru.\arabic{equation}}
\begin{center}
{\small\bfseries RUSSIAN TEXT: RETYPESET TRANSCRIPTION\par}
\end{center}
\noindent{\small This appendix reproduces the Russian transcription used
for the translation. It is not a facsimile of the 1941 printing.}
\begin{russian}
\sloppy
\begin{center}
{\bfseries В. Н. ЗАСУХИН}\par
\vspace{0.3em}
{\bfseries К ТЕОРИИ МНОГОМЕРНЫХ СТАЦИОНАРНЫХ СЛУЧАЙНЫХ\\
ПРОЦЕССОВ}\par
\vspace{0.6em}
{\itshape (Представлено академиком А. Н. Колмогоровым 10 XI 1941)}
\end{center}

\origpage{435}
Пусть каждому целому $t(-\infty<t<+\infty)$ и каждому $k=1,2,\ldots,n$
поставлена в соответствие комплексная случайная величина $x_k(t)$.
Совокупность случайных величин $x_k(t)$ образует $n$-мерный дискретный
стационарный случайный процесс в смысле H. Cramér'a (см. \refsup{1}),
если математические ожидания
\[
B_{ij}(m)=\M\bigl[x_i(t+m)\overline{x_j(t)}\bigr]
\]
конечны и не зависят от $t$ (принимаемое в работе H. Cramér'a дополнительное
допущение $\M[x_i(t)]=0$ для дальнейшего не существенно). В дальнейшем
такого рода совокупность случайных величин будет называться просто
$n$-мерным процессом и обозначаться $\{x(t)\}$.

Как указано в работах А. Н. Колмогорова \refsup{2} и \refsup{3}, при изучении
стационарных случайных процессов естественно пользоваться геометрией
унитарного пространства $\mathfrak S$ всех комплексных случайных величин $z$
данной задачи, имеющих конечное математическое ожидание $\M|z|^2$. Обозначим,
следуя А. Н. Колмогорову, через $H_x$ минимальное линейное подпространство
пространства $\mathfrak S$, содержащее все элементы
$x_k(t)(-\infty<t<+\infty,\ k=1,2,\ldots,n)$, и через $H_x(t)$, минимальное
линейное подпространство пространства $\mathfrak S$, содержащее все $x_k(s)$,
для которых $s\le t$.

\noindent\textbf{Определение 1.} \emph{Процесс $\{x(t)\}$ называется сингулярным, если
$H_x(t)=H_x$ при любом $t$.}

Сингулярные процессы совпадают с процессами, допускающими сколь угодно
точный прогноз их будущего течения (значений $x_k(t)$ для $t>t_0$) по их
течению в прошлом (значениям $x_k(t)$ для $t\le t_0$) при помощи линейных
формул прогноза с постоянными коэффициентами (см. по этому поводу работу
А. Н. Колмогорова \refsup{2}).

\noindent\textbf{Определение 2.} \emph{Процесс $\{x(t)\}$ называется регулярным, если
пересечение $S_x$ пространств $H_x(t)$, соответствующих всем целым значениям
$t$, состоит из единственной точки 0, но не все $x_k(t)$ равны нулю.}

Следующая теорема показывает, что произвольный не сингулярный

\origpage{436}

процесс может
быть вполне определенным образом представлен в виде суммы сингулярного и
регулярного процессов:

\noindent\textbf{Теорема 1.} \emph{Каждый $n$-мерный не сингулярный процесс
$\{x(t)\}$ может быть представлен в виде}
\begin{equation}
 x_k(t)=s_k(t)+r_k(t),
\end{equation}
\emph{где $n$-мерные процессы $\{s_k(t)\}$ и $\{r_k(t)\}$ обладают следующими
свойствами:}

\noindent\emph{1) процесс $\{s_k(t)\}$ сингулярен,}

\noindent\emph{2) процесс $\{r_k(t)\}$ регулярен,}

\noindent\emph{3) $\M[s_i(s)\overline{r_j(t)}]=0$}\par\nopagebreak[4]

\noindent\emph{при любых $i,j,s$ и $t$}

\noindent\emph{4) при любых $k$ и $t$ элементы $s_k(t)$ и $r_k(t)$ принадлежат
пространству $H_x$.}

\emph{Представление (1), удовлетворяющее условиям 1)--4), единственно.}

Процессы $\{s(t)\}$ и $\{r(t)\}$, удовлетворяющие условиям теоремы 1,
называются, соответственно, сингулярной и регулярной компонентой процесса
$\{x(t)\}$. Элементы $s_k(t)$ сингулярной компоненты суть не что иное, как
ортогональные проекции соответствующих элементов $x_k(t)$ в пространство
$S_x$. Очевидно, что регулярный процесс совпадает со своей регулярной
компонентой. Природа регулярных процессов в значительной мере характеризуется
следующей теоремой:

\noindent\textbf{Теорема 2.} \emph{Каждый $n$-мерный регулярный процесс $\{r(t)\}$
может быть представлен в виде}
\begin{equation}
 r_k(t)=\sum_{m=0}^{\infty}\sum_{j=1}^{j=q}c_{kj}(m)u_j(t-m),
\end{equation}
\emph{где:}

\noindent\emph{1) $q\le n$;}

\noindent\emph{2) комплексные коэффициенты $c_{kj}(m)$ таковы, что матрица
$\|c_{kj}(0)\|$ имеет ранг $q$;}

\noindent\emph{3) случайные величины $u_k(t)$ $(-\infty<t<+\infty,\ k=1,2,\ldots,q)$
образуют $q$-мерный процесс, для которого}
\[
\M|u_k(t)|^2=1
\]
\emph{при любых $k$ и $t$ и}
\[
\M\bigl[u_i(s)\overline{u_j(t)}\bigr]=0,
\]
\emph{когда $i\ne j$, или $s\ne t$.}

\noindent\emph{4) элементы $u_k(t)$ принадлежат соответствующим пространствам
$H_x(t)$.}

Во всех представлениях вида (2) данного регулярного процесса,
удовлетворяющих условиям 1)--4), $q$ имеет одно и то же значение. Любые два
таких представления связаны соотношениями
\[
 u'_i(t)=\sum_{j=1}^{j=q}a_{ij}u_j(t),
\]
где коэффициенты $a_{ij}$ не зависят от $t$.

Число $q$ называется рангом регулярного процесса $\{r(t)\}$. Рангом
произвольного не сингулярного процесса называется ранг его регулярной
компоненты. Сингулярный процесс имеет по определению

\origpage{437}

ранг равный нулю.
Представления (1) и (2) являются обобщением на $n$-мерные процессы основного
результата, полученного H. Wold'ом при изучении одномерных процессов (см.
книгу H. Wold'a\refsup{4}, или теорему 18 из \refsup{3}). Доказательство
сформулированных выше теорем 1 и 2 не представляет особых затруднений, если
пользоваться методами из \refsup{3}.

Более трудными представляются вопросы, относящиеся к установлению спектральных
характеристик различных типов стационарных процессов. Автору удалось получить
в этом направлении следующий результат, обобщающий теорему 22 из \refsup{3}:

\newpage
\noindent\textbf{Теорема 3.} \emph{Для того, чтобы $n$-мерный процесс $\{x(t)\}$ был
регулярным ранга $n$, необходимо и достаточно соблюдение совокупности трех
условий:}

\noindent\emph{1) все спектральные функции $F_{ij}(\lambda)$ процесса $\{x(t)\}$
должны быть абсолютно непрерывны;}

\noindent\emph{2) детерминант}
\[
\Delta(\lambda)=\bigl|F'_{ij}(\lambda)\bigr|
\]
\emph{из производных от спектральных функций должен быть почти всюду
положителен;}

\noindent\emph{3) интеграл}
\[
J=\int_{-\pi}^{+\pi}\log\Delta(\lambda)\,d\lambda
\]
\emph{должен быть конечен.}

При этом, спектральные функции $F_{ij}(\lambda)$ определяются в соответствии с
\refsup{3} формулами
\[
B_{ij}(m)=\int_{-\pi}^{+\pi}e^{im\lambda}\,dF_{ij}(\lambda),
\qquad
F_{ij}(-\pi)=0.
\]

\vspace{0.6em}
\noindent
\begin{minipage}[t]{0.60\textwidth}
Математический Институт\\
Московского Государственного Университета
\end{minipage}%
\begin{minipage}[t]{0.38\textwidth}
\raggedleft
Поступило\\
10 XI 1941
\end{minipage}

\vspace{1.5em}
\begin{center}
\textbf{ЦИТИРОВАННАЯ ЛИТЕРАТУРА}
\end{center}

\noindent\textsuperscript{1} H. Cramér, Annals of Mathematics, 41, 215--230 (1940).
\textsuperscript{2} А. Н. Колмогоров, Известия АН СССР, Серия математическая,
5, 3--14 (1941). \textsuperscript{3} А. Н. Колмогоров, Бюллетень Московского
Государственного Университета, Математика, т. 2, вып. 6, (1941).
\textsuperscript{4} H. Wold, A study in the analysis of stationary time series,
Uppsala 1938.

\end{russian}
\end{document}